\documentclass[9pt,twocolumn,twoside]{pnas-new}
\usepackage{mathabx}
\usepackage{xcolor}
\usepackage{array}
\templatetype{pnasresearcharticle}

\title{Bacteria push the limits of chemotactic precision to navigate dynamic chemical gradients}

\author[a,1,2]{Douglas R. Brumley}
\author[b,1,2]{Francesco Carrara} 
\author[c]{Andrew M. Hein}
\author[d,e]{Yutaka Yawata}
\author[f]{Simon A. Levin}
\author[b,2]{Roman Stocker}

\affil[a]{School of Mathematics and Statistics, The University of Melbourne, Parkville, Victoria 3010, Australia}
\affil[b]{Institute of Environmental Engineering, Department of Civil, Environmental and Geomatic Engineering, ETH Zurich, 8093 Zurich, Switzerland}
\affil[c]{Institute of Marine Sciences, University of California, Santa Cruz, CA 95060, USA}
\affil[d]{Faculty of Life and Environmental Sciences, University of Tsukuba, Tsukuba, Ibaraki 305-8572, Japan}
\affil[e]{Microbiology Research Center for Sustainability, University of Tsukuba, Tsukuba, Ibaraki 305-8572, Japan}
\affil[f]{Department of Ecology and Evolutionary Biology, Princeton University, Princeton, NJ 08544, USA}
\leadauthor{Brumley}

\significancestatement{The limited precision of sensory organs places fundamental constraints on organismal performance. An open question, however, is whether organisms are routinely pushed to these limits, and how limits might influence interactions between populations of organisms and their environment. By combining a method to generate dynamic, replicable resource landscapes, high-speed tracking of freely moving bacteria, a new mathematical theory, and agent-based simulations, we show that sensory noise ultimately limits when and where bacteria can detect and climb chemical gradients. Our results suggest the typical chemical landscapes bacteria inhabit are dominated by noise that masks shallow gradients, and that the spatiotemporal dynamics of bacterial aggregations can be predicted by mapping the region where gradient signal rises above noise.}

\authorcontributions{Author contributions: D.R.B., F.C., A.M.H., Y.Y., S.A.L., and R.S. designed research; D.R.B., F.C., and A.M.H. analyzed data; D.R.B., F.C., A.M.H., and R.S. wrote the paper; D.R.B. performed numerical simulations; D.R.B., F.C., and A.M.H. developed theory; and F.C.performed experiments.}
\authordeclaration{The authors declare no conflict of interest.}
\equalauthors{\textsuperscript{1}D.R.B. and F.C. contributed equally to this work.}
\correspondingauthor{\textsuperscript{2}To whom correspondence may be addressed. E-mail: d.brumley@unimelb.edu.au, carraraf@ethz.ch, or romanstocker@ethz.ch.}

\keywords{chemotaxis $|$ motility $|$ sensing noise $|$ microbial ecology $|$ ocean} 
\begin{abstract}
Ephemeral aggregations of bacteria are ubiquitous in the environment, where they serve as hotbeds of metabolic activity, nutrient cycling, and horizontal gene transfer. In many cases, these regions of high bacterial concentration are thought to form when motile cells use chemotaxis to navigate to chemical hotspots. However, what governs the dynamics of bacterial aggregations is unclear. Here, we use a novel experimental platform to create realistic sub-millimeter scale nutrient pulses with controlled nutrient concentrations. By combining experiments, mathematical theory and agent-based simulations, we show that individual \textit{Vibrio ordalii} bacteria begin chemotaxis toward hotspots of dissolved organic matter (DOM) when the magnitude of the chemical gradient rises sufficiently far above the sensory noise that is generated by stochastic encounters with chemoattractant molecules. Each DOM hotspot is surrounded by a dynamic ring of chemotaxing cells, which congregate in regions of high DOM concentration before dispersing as DOM diffuses and gradients become too noisy for cells to respond to. We demonstrate that \textit{V. ordalii} operates close to the theoretical limits on chemotactic precision. Numerical simulations of chemotactic bacteria, in which molecule counting noise is explicitly taken into account, point at a tradeoff between nutrient acquisition and the cost of chemotactic precision. More generally, our results illustrate how limits on sensory precision can be used to understand the location, spatial extent, and lifespan of bacterial behavioral responses in ecologically relevant environments. 
\end{abstract}

\doi{\url{www.pnas.org/cgi/doi/10.1073/pnas.1816621116}}
\begin{document}

\maketitle
\thispagestyle{firststyle}
\ifthenelse{\boolean{shortarticle}}{\ifthenelse{\boolean{singlecolumn}}{\abscontentformatted}{\abscontent}}{}

\dropcap{M}otile bacteria often survive by consuming ephemeral sources of dissolved organic matter (DOM) produced, for example, in the ocean by phytoplankton lysis and exudation, or sloppy feeding and excretion by larger organisms \cite{Fuhrman1987, Blackburn1998, Stocker2012_MMBR, Jackson2012}. The microscale interactions between nutrient sources and bacteria underpin ocean biogeochemistry and are strongly influenced by the ability of bacteria to actively navigate towards favorable conditions. Past experiments on chemotaxis using {\it Escherichia coli} and other model bacteria have generally focused on stable gradients of intermediate to high nutrient concentrations, where bacteria can readily detect chemical gradients \cite{Segall1986, Brown1974, Waite2016}. However, the environments that wild bacteria navigate are often characterized by short-lived, microscale chemical gradients where background conditions are highly dilute \cite{Stocker2012, Smriga2016}. In such ephemeral chemical fields, bacteria experience a gradient in DOM concentration as a noisy, dynamic signal, rather than as a steady concentration ramp \cite{Hein2016_PNAS}.

Chemotactic bacteria rely on temporal gradient sensing to bias their swimming behavior according to whether their measurement of a chemical concentration is rising or falling over time. Such measurements are accomplished using sophisticated receptors on their surface, combined with intracellular transduction pathways \cite{Wadhams2004}. This process fundamentally involves interaction with discrete chemoattractant molecules \cite{BergPurcell1977}: intrinsic randomness in the encounter rate affects a cell's measurement of the gradient. This randomness places fundamental constraints on the cell's ability to resolve gradients.

Theoretically, the relationship between the magnitude of a gradient signal and the noise associated with a cell's measurement of that signal -- the signal to noise ratio (SNR) -- determines when and where cells can perform chemotaxis. Recent theoretical work has explored the physical limits on the accuracy and precision of cellular gradient sensing \cite{Endres2006, Mora2010}, expanding on the seminal work of Berg and Purcell \cite{BergPurcell1977}.

In natural environments, gradients are often noisy, in part due to low concentrations and local fluctuations, and can change over timescales comparable to the chemotactic response \cite{Frankel2014, Hein2016, Taylor2012}. Understanding what governs chemotaxis and aggregation of bacteria in these noisy, ephemeral environments requires coupling an experimental method for precisely quantifying bacterial responses to microscale nutrient pulses, with a theoretical framework that specifically incorporates sensory noise. This has so far remained elusive.

\vspace{-5pt}
\subsection*{Quantifying Chemotaxis in Realistic Microenvironments}  

\begin{figure}[t!]
\begin{center}
\includegraphics[width=\columnwidth]{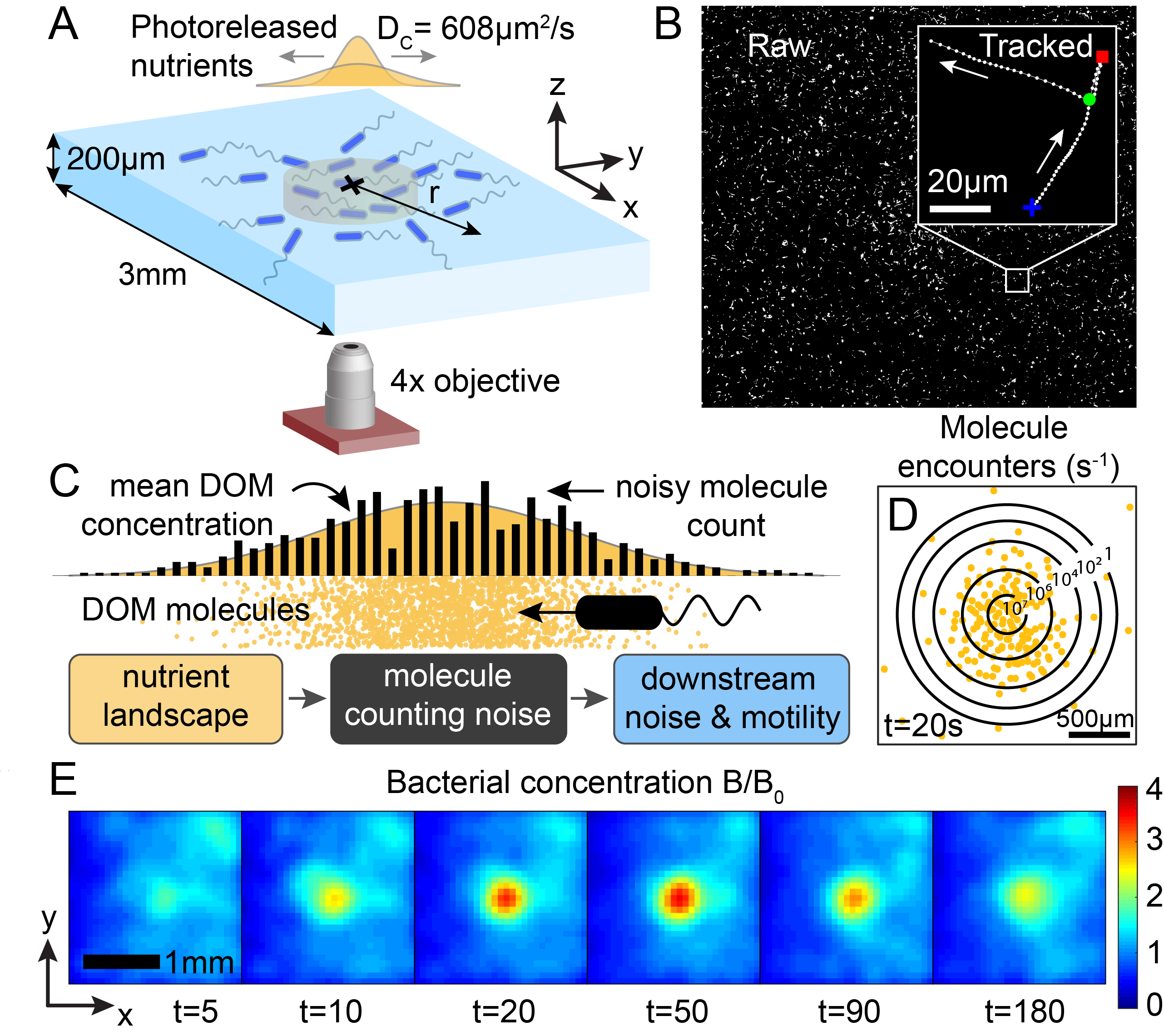}
\caption{Bacteria are able to perform chemotaxis in the presence of molecule counting noise. (A) Localized chemoattractant becomes available at the center of the chamber through photorelease of caged glutamate (orange), which subsequently diffuses and attracts chemotactic bacteria (blue). (B) Bacterial trajectories are extracted across the domain, revealing runs (white), reversals (red) and flicks (green). (C) The discrete nature of glutamate encounters introduces noise into the bacterium's gradient measurement on top of other sources of noise. (D) Contours showing the instantaneous rate of encounter with glutamate molecules experienced by bacteria (shown at $t=20$\,s). (E) Despite the noise, the cells exhibit strong accumulation with bacterial cell concentration, $B/B_0$, reaching a high value over the background a few tens of seconds following the pulse release. Positions of individual bacteria have been binned into 75\,$\mu$m $\times$ 75\,$\mu$m grids and averaged over 8-10\,s intervals. \vspace{-15pt}}
\label{FIG1}
\end{center}
\end{figure}

To create controlled, dynamic nutrient pulses that mimic those bacteria interact with in the ocean, we developed a system to introduce and make almost instantly available to the bacteria an amino acid source with known concentration into a chemically dilute background within a microfluidic chamber \cite{Son2015} (SI Appendix). Prior to the experiment, the chamber is filled with a known concentration of 4-methoxy-7-nitroindolinyl-caged-L-glutamate, a `caged' version of the amino acid glutamate -- a potent chemoattractant and one of the most abundant dissolved amino acids in coastal waters \cite{Dawson2011}. When bound to the cage, glutamate was undetectable by the bacteria. By exposing the center of the chamber to a focused LED pulse, a controlled quantity of glutamate is photoreleased \cite{Jikeli2015, McCray1989} in a vertical column (Fig.~\ref{FIG1}A). The amount of glutamate can be controlled to match the amino acids released from a lysing phytoplankton cell \cite{Blackburn1998}. In the experiments, this is varied in the range 0.0088--0.22 pmol, where the number of molecules released in the pulse can be determined by a calibration relationship between exposure time and uncaging fraction (see Fig.~\ref{FIG3} and SI Appendix, Fig. S4). The subsequent diffusion of this axisymmetric cylindrical pulse (diffusivity $D_C=608\,\mu \text{m}^2 \, \text{s}^{-1}$) is well-approximated by a point source spreading with a Gaussian profile $C(r,t)$ (see SI Appendix, Fig.~S5). The instantaneous rate at which bacteria of radius $a$ encounter glutamate molecules is $R = 4 \pi a D_C N_A C(r,t)$ where $N_A$ is the Avogadro constant. At $t=20\,$s after pulse release, a considerable number of bacteria in the domain encounter just a few molecules per second (Fig.~\ref{FIG1}D), highlighting the importance of considering the discrete nature of the chemoattractant.

\begin{figure}[tbhp]
\begin{center}
\includegraphics[width=\columnwidth]{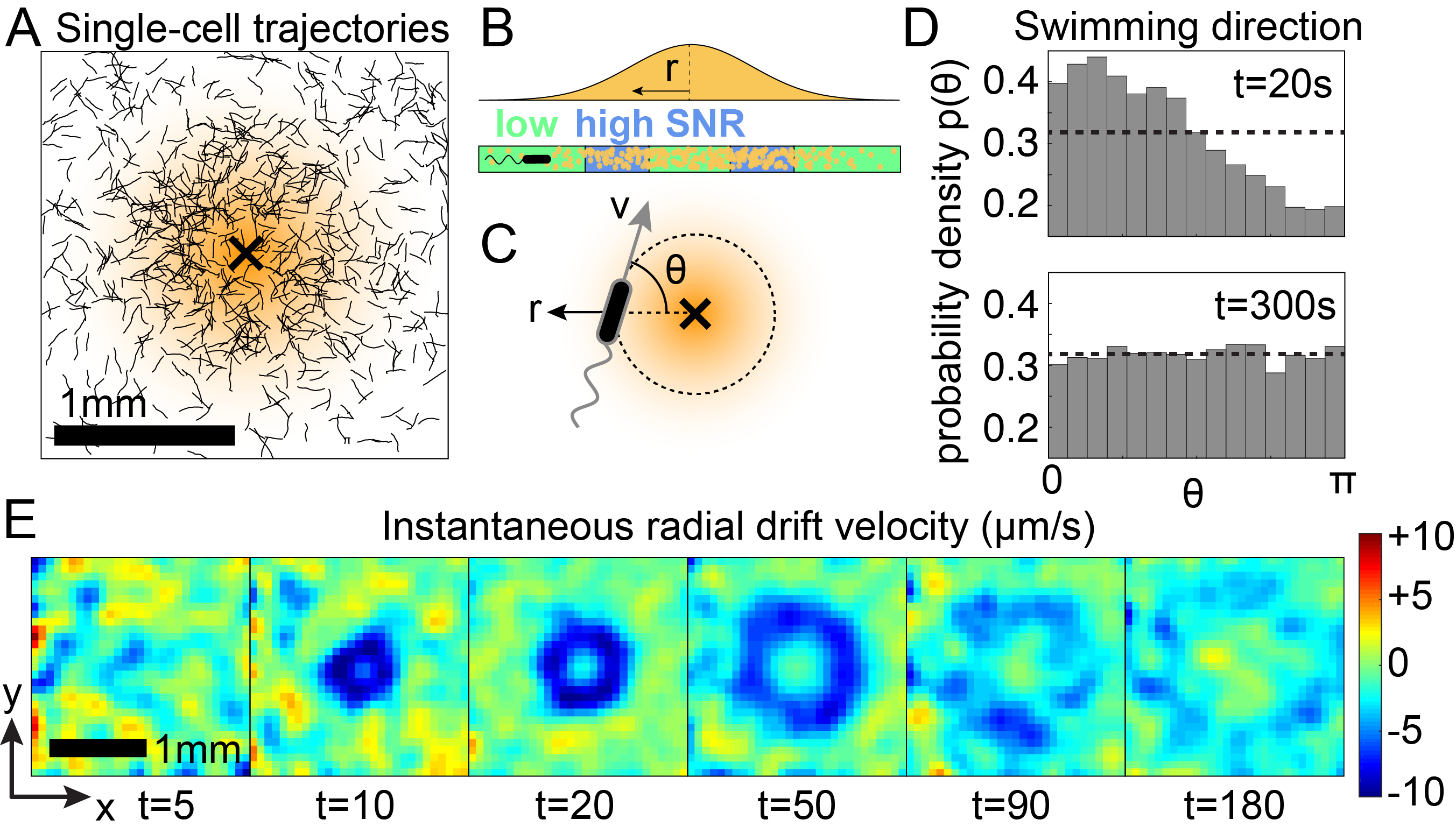}
\caption{Bacterial drift is confined to discrete regions in space and time where the signal-to-noise ratio is sufficiently high. (A) Single-cell trajectories in the first 60\,s following pulse release. (B) Schematic of the mean glutamate concentration as a function of the distance from the center of a pulse. Far from and near to the pulse center, the signal-to-noise ratio (SNR) is low. At an intermediate distance, concentration gradients are strong and the gradient signal can emerge above noise. (C) The swimming velocity of each bacterium makes an angle $\theta$ relative to the radial vector towards the pulse center. (D) Probability distribution, $p(\theta)$, of the swimming angle $\theta$ within the region $r \in$ [300$\mu$m,400$\mu$m]. The dotted lines correspond to isotropic distributions (random swimming direction). A strong inward bias, $E[\theta] < \pi/2$, is observed at $t=20$\,s. (E) Instantaneous radial drift velocity, $v_{\text{drift}}$, at various times following the pulse release. \vspace{-15pt}}
\label{FIG2}
\end{center}
\end{figure}

To measure how cells respond to this rarefied chemical pulse (Fig.~\ref{FIG1}D), we recorded over 1 million bacterial trajectories (Fig.~\ref{FIG1}B), over 20 min starting at pulse release, for three replicate pulses. This allowed us to measure motion at the single-cell level and to quantify chemotactic behavior at the population level. For each track (Fig.~\ref{FIG2}A) we quantified the angle, $\theta$, between the cell's instantaneous swimming velocity and the vector pointing from the cell position to the center of the pulse (Fig.~\ref{FIG2}C). At $t=20$\,s after pulse release, there is a strong inward bias ($0 < E[\theta]  < \pi/2$) for bacteria in the region $r \in$ [300$\mu$m, 400$\mu$m], whereas at either larger radii or later times ($t>$300\,s) the swimming is isotropic (Fig.~\ref{FIG2}D). Figure~\ref{FIG2}E shows spatial snapshots of the average radial velocity $v_{\text{drift}} = \langle - v \cos \theta \rangle$ at various times after the pulse release. An annular region of biased bacterial motion with $v_{\text{drift}}<0$ expands around the pulse center and eventually disappears. The chemotactic response of the bacteria gives rise to the transient accumulation of cells near the center, with concentrations up to four times higher than the background bacterial concentration $B_0$ within 1 min of pulse release (Fig.~\ref{FIG1}E). In what follows, we show that the region of chemotaxis and the rate at which cells move toward the pulse depend not only on the glutamate gradient, but also on the noise associated with cellular measurements of that gradient.

\subsection*{Measurement Noise and Bacterial Chemotaxis}

Small motile cells in the size range of {\it E. coli}, sperm cells \cite{Kromer2018}, many {\it Vibrio}s, and other bacteria cope with high levels of noise when estimating the concentrations and gradients of chemicals \cite{Mora2010, Hein2016}. Over some short time interval $(t_0,t_0+T)$ the true local concentration at the cell's position, $\mathbf{x}$, is (SI Appendix) $c(t) \approx c_0 + c_1(t-t_0) = c_0 + \big[\partial c / \partial t + \mathbf{v} \cdot \nabla c \big] (t-t_0)$, where $\mathbf{v} = \dot{\mathbf{x}}$. Mora and Wingreen \cite{Mora2010} showed that there is a bound on the precision with which a cell can estimate the gradient component of this local concentration. A cell that seeks to estimate the concentration gradient over a time interval $T$ is limited by the following performance bound:
\begin{equation}
\sigma_{\hat{c}_1} \geq \sigma^{\text{min}}, \quad \text{where} \quad  \sigma^{\text{min}}  = \bigg[\frac{3 c_0}{\pi a D_C T^3} \bigg]^{1/2}, \label{minstd}
\end{equation}
and $\sigma_{\hat{c}_1}$ is the standard deviation (SD) of the cell's estimate of the gradient, $D_C$ is the diffusivity of the chemical compound, $a$ is the radius of the bacterial cell, and $c_0$ is the local average concentration of the compound \cite{Mora2010,Hein2016}. The limit defined by \eqref{minstd} assumes that the only source of uncertainty in the cell's gradient estimate is molecule counting noise, introduced by stochastic encounters between the cell and chemoattractant molecules \cite{Mora2010,Hein2016}. Equation~(\ref{minstd}) is derived by idealizing a bacterium as a perfectly absorbing sphere \cite{BergPurcell1977} and does not require assumptions about the species' biology or behavior.

Equation~(\ref{minstd}) illustrates that the uncertainty in the gradient measurement depends on the local chemical concentration, $c_0$. Therefore, the ability of a bacterium to accurately estimate and climb the gradient in our experiment depends strongly on where the cell is located (Fig.~\ref{FIG2}B). As an example, shortly after pulse release, a cell $\sim$1 mm from the pulse center in the dilute conditions of our experiment would experience fluctuations in molecular encounters which preclude measurement of the chemical gradient within this timescale. Likewise, near the center of the pulse, the gradient is weak (Fig.~\ref{FIG2}B) and the measurement noise will dominate the gradient signal \cite{Hein2016}. The ring-shaped region where the gradient is high and noise is moderate (Fig.~\ref{FIG2}B) suggests that the relationship between gradient signal and sensory noise is responsible for the annular region of chemotaxis observed in our experiment (Fig.~\ref{FIG2}E). We explored this hypothesis using a computational model of chemotaxis that includes measurement noise.

\subsection*{Linking Measurement Noise and Chemotactic Performance}

To determine whether and how measurement noise affected the chemotactic response of bacteria in our experiments, we developed a simplified model of the \textit{Vibrio} chemotaxis response that incorporates the essential features of bacterial navigation. Many details of the chemotaxis pathway that are known for \textit{E. coli} \cite{Cluzel2000, Sourjik2002, Berg2004} are not known for \textit{V. ordalii}, nor are such details known for most non-model bacteria. We therefore modeled \textit{Vibrio} chemotaxis using a minimal model inspired by Long {\it et al.} \cite{Long2017} to combine the physical theory of Eq.~(\ref{minstd}) with the essential features of gradient measurement, adaptation, and motor output. For each bacterium we model an internal state variable, $S(t)$, which evolves according to (SI Appendix):
\begin{equation}
\dot{S}(t) = - \frac{S}{t_M} + \kappa \mathcal{M}(\mathbf{x},\mathbf{v},t), \label{chemotaxis_model}
\end{equation}
where $S=0$ is the adapted state, $t_M$ is the adaptation timescale associated with methylation dynamics ($\sim$1.3 s in \textit{Vibrio} \cite{Xie2015}), $\kappa$ is the receptor gain rescaled by the half-saturation constant, and $\mathcal{M}(\mathbf{x},\mathbf{v},t)$ is the (noisy) concentration gradient perceived by the cell, which is subject to the bounds of Eq.~(\ref{minstd}). Within each time interval of duration $T$, we model the gradient estimate as a normally distributed random variable, $\mathcal{M}(\mathbf{x},\mathbf{v},t) = \mathcal{N}(\mu, \sigma^2)$, with mean $\mu = (\partial/\partial t + \mathbf{v} \cdot \nabla ) C(\mathbf{x},t)$ and SD $\sigma = \Pi \times \sigma^{\text{min}} = \Pi \big[ 3 C(\mathbf{x},t) / \pi a D_C T^3 \big]^{1/2}$. Here, we have assumed that the SD of the cell's estimate of the true ramp rate is proportional to the theoretical bound given in Eq.~(\ref{minstd}) with a proportionality constant equal to $\Pi \geq 1$. We will refer to $\Pi$ as a ``precision factor'' because it expresses how precise a cell's estimate of the gradient is relative to the theoretical bound. $\Pi = 1$ means the cell has reached the bound, $\Pi = 10$ means the cell is ten times less precise than the bound, and so forth. This formulation assumes that noise in the dynamics of the internal variable can be captured as a multiple of the lower bound on noise in the gradient measurement itself. This can be interpreted as additional gradient measurement noise above the limit set by Eq.~(\ref{minstd}) (e.g. as a result of sub-optimal measurement or transduction \cite{Mora2010}) or as noise introduced later in the transduction pathway. We do not attempt to distinguish these possibilities.

Equation (\ref{chemotaxis_model}) involves two distinct timescales: the adaptation timescale $t_M$ and the gradient estimate timescale $T$. While $T$ has not been directly measured, it is bounded by measurable features of the chemotaxis response. Firstly, it cannot be shorter than the typical phosphorylation timescale, since transduction of the receptor binding kinetics is a necessary precursor to processing and integrating this information. This sets a lower bound of $T \sim$ 100 ms \cite{Morton-Firth1999,Sourjik2002}. Secondly, $T$ must be less than the run time, $\tau$, of the cell if the cell is to consistently respond to a spatial gradient \cite{BergPurcell1977}. Our data indicate a sharp cutoff to the distribution of run times at 120 ms (SI Appendix, Fig.~S3C). This suggests that $T$ is of the order $0.1$ s, and we will assume this value in our model.
	
\textit{V. ordalii} exhibits run-reverse-flick locomotion (Fig.~\ref{FIG1}B). We simulate its chemotactic behavior by modeling transitions from run to reverse and from reverse to flick, assuming the associated switching rates to be governed by a nonhomogeneous Poisson process with rate $\lambda(S)$ \cite{Long2017} (SI Appendix). This formulation allows the cell to modulate its mean run time, $\tau(S) = 2 \tau_0 / (1+\exp (-\Gamma S))$, from the unbiased value $\tau_0$ (experimentally determined with no chemical gradient), where $\Gamma$ is a dimensionless motor gain.

\begin{figure}[t!]
\begin{center}
\includegraphics[width=\columnwidth]{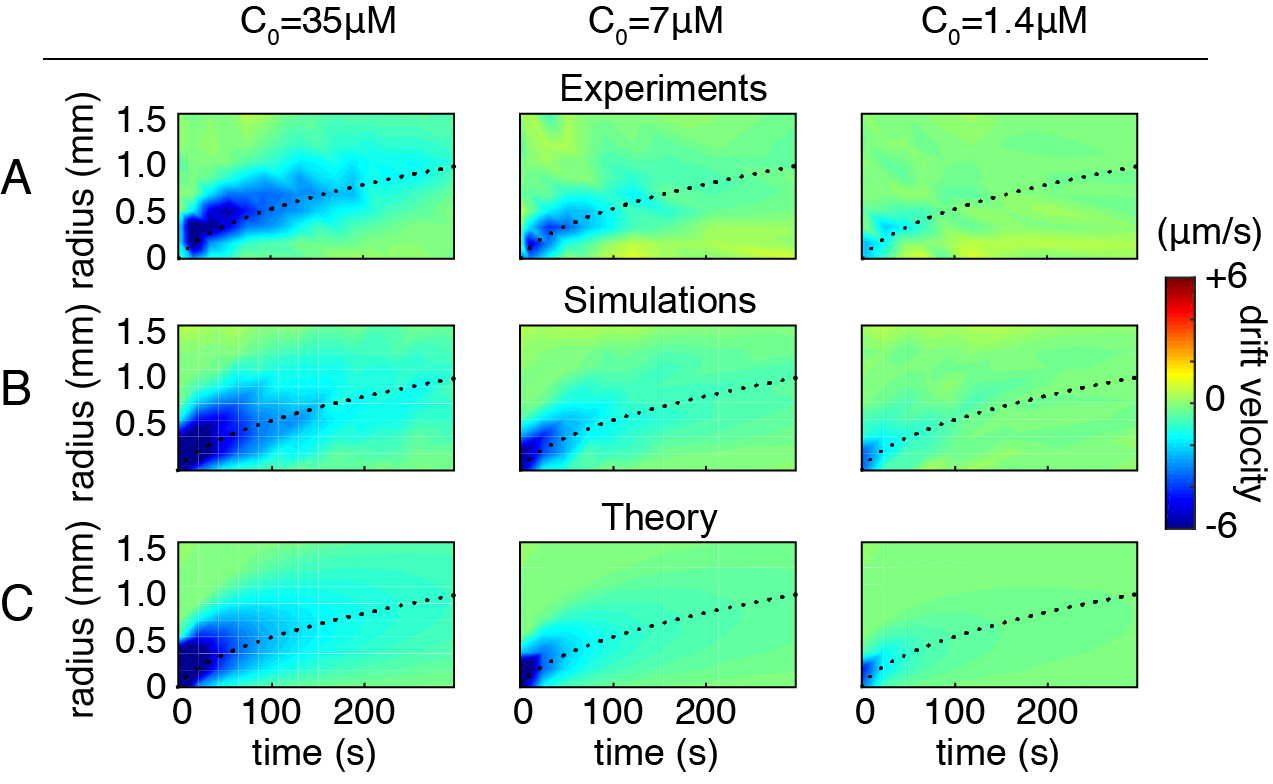}
\caption{Simulations and theory which each incorporate molecule counting noise successfully predict zones of chemotaxis across various pulse sizes. (A) Experimentally measured radial drift velocity as a function of time and distance from the pulse, for initial pulses of concentration $C_0=35$, 7, and 1.4 $\mu$M. (B) Agent-based simulations and (C) theoretical predictions with fitted precision factor and gain. \vspace{-15pt}}
\label{FIG3}
\end{center}
\end{figure}

\vspace{-5pt}
\subsection*{Predicting Chemotactic Performance of a Population} Using the above model of chemotaxis, we performed 3D agent-based simulations of populations of bacteria foraging in the dynamic nutrient landscape studied in our experiments. Cells are subject to rotational diffusion, and execute run-reverse-flick motion (Fig.~\ref{FIG1}B inset) with reorientation angles drawn from distributions for a closely related {\it Vibrio} species \cite{Son2013} (SI Appendix). The agent-based model was compared to experimental data by fitting the precision factor $\Pi$, the motor gain $\Gamma$, and the rescaled receptor gain $\kappa$ to data on the radial drift velocity of bacteria from our experiment (SI Appendix). The results depend more strongly on the precision factor than on either of the gains (Fig.~S6). The spatiotemporal evolution of the drift velocity from our experiments (Fig.~\ref{FIG3}A) is captured by the computational model (Fig.~\ref{FIG3}B), with the formation of an expanding -- and eventually disappearing -- annular region of chemotaxing cells. Outside this dynamic annulus, the shallow gradients are masked by noise (Fig. \ref{FIG2}B) and bacterial motion is unbiased. These results involving drift velocity are not to be confused with the previously reported ``volcano effect'' in bacterial density \cite{Bray2007}.

\begin{figure}[t!]
\begin{center}
\includegraphics[width=\columnwidth]{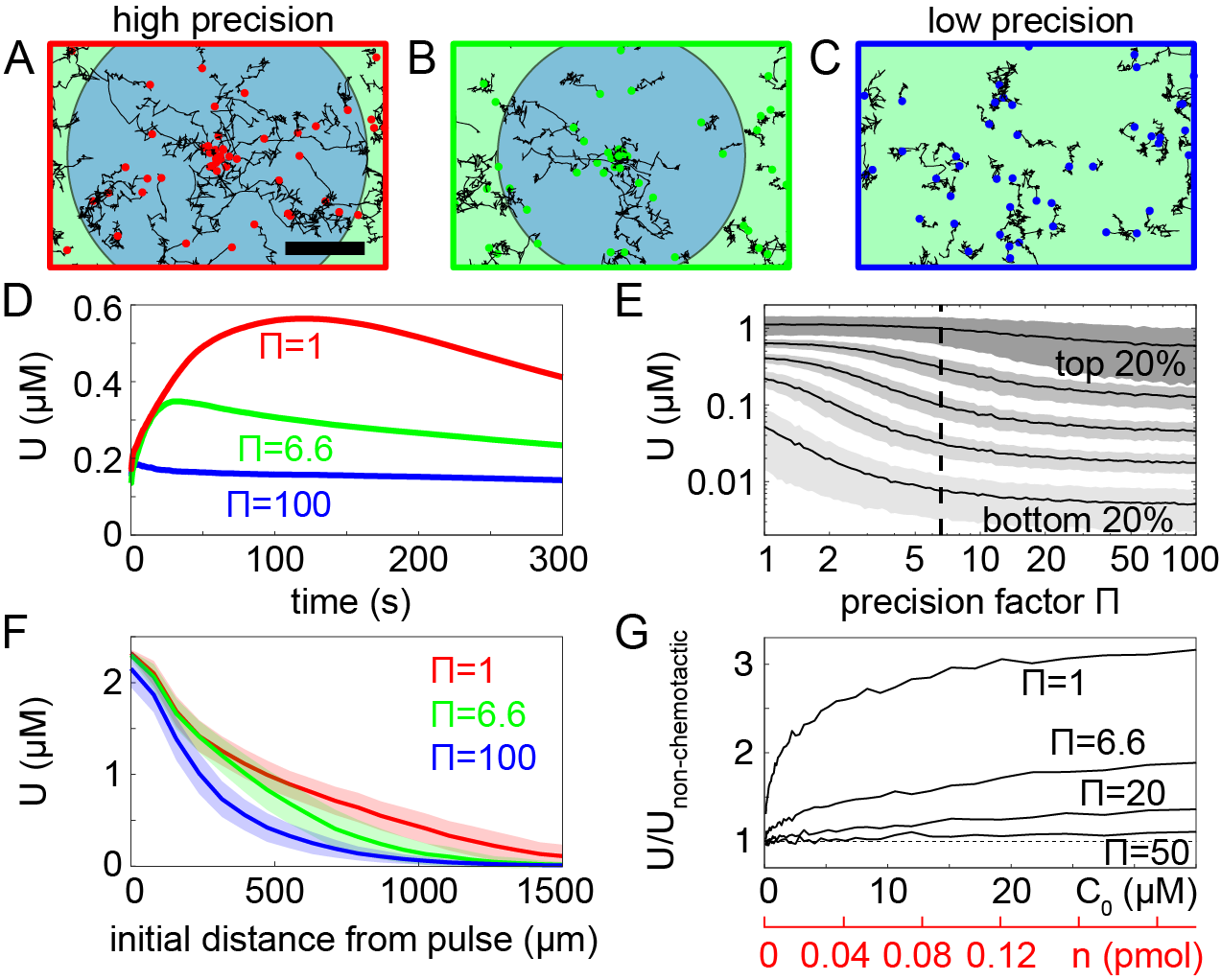}
\caption{(A-C) Numerical simulations of bacteria for varying values of the chemotactic precision factor ($\Pi=1$, 6.6, 100) show that aggregation depends strongly on chemotactic precision. Sample trajectories of 100 model bacteria followed for 60~s are shown (see videos S1-S3). The shaded circular zones (blue) represent a measured drift velocity $|v_{\text{drift}}| > 1 \ \mu$m/s over the first 60\,s. The black scale bar represents 500\,$\mu$m. (D) Nutrient exposure $U$ as a function of time, averaged across the population. Results are shown for three different precision factors, $\Pi$, corresponding to those in panels A-C. (E) Time-averaged nutrient exposure as a function of chemotactic precision for different quintiles of the bacterial population. (F) Mean nutrient exposure (over 300\,s) plotted in terms of the initial distance of bacteria from the pulse. Shaded regions show standard error. Panels D-F correspond to $C_0 = 35 \, \mu$M. (G) Enhancement in potential uptake compared to non-chemotactic cells as a function of pulse size. \vspace{-15pt}}
\label{FIG4}
\end{center}
\end{figure}

The agreement between model and data was good across nutrient pulses of different intensities. We performed additional experiments in which we decreased the concentration of uncaged glutamate in the pulse by a factor of 5 and a factor of 25 (corresponding to $C_0=7 \, \mu$M and $C_0=1.4 \, \mu$M, respectively). The model accurately predicts how chemotaxis varies with pulse size (Fig.~\ref{FIG3}B). A single value of the precision factor, $\Pi=6.6$, captures the drift velocity profile across all pulse sizes with mean fitting error $\sim$1\% of the swimming speed.

The observed value of $\Pi=6.6$ is relatively close to the theoretical bound of $\Pi=1$ and is within the same order of magnitude of rough estimates of the chemotactic precision of {\it E. coli} \cite{BergPurcell1977, Bialek2005}. This is remarkable, particularly when one considers that bacterial behavior is influenced not only by molecule counting noise, but also by internal sources of noise associated with the signaling network \cite{Korobkova2004, Keegstra2017}, receptor readout copies \cite{Govern2014}, and other cellular functions (e.g., running a flagellar motor) \cite{Bialek2005}. Similar values of precision are estimated when nutrient consumption is included ($\Pi=4.2$), adaptation is neglected ($\Pi=8.4$), or both ($\Pi=4.4$) (SI Appendix, Fig.~S8), highlighting the robustness of the finding that {\it V. ordalii} appears to perform close to the theoretical limit imposed by gradient measurement.

It will be important in future studies of the chemotaxis network to determine precisely the biochemical processes that set $T$. However, based on the timescale set by the phosphorylation and dephosphorylation cycle of CheY in {\it E. coli} \cite{Govern2014}, we expect the value of $T$ to be very close to 0.1 s. Although the estimated value of the precision factor, $\Pi$, depends on the value of $T$, the dependence is not as strong as it might appear from \eqref{minstd}. For example, setting $T=0.05 \,$s or $T=0.2 \,$s and re-estimating $\Pi$ from the simulations results in values of $\Pi=3$ and $\Pi=13$, respectively. This weaker dependence of $\Pi$ on $T$ is due to partial compensation by the fitted gain term.

\vspace{-5pt}
\subsection*{Continuum Theory}

In addition to the computational model described, we developed a simplified continuum model to predict the chemotactic drift velocity, that incorporates gradient signal and measurement noise over a single run-reverse cycle (SI Appendix). Consider a cell that makes a run with speed $v$ up a (locally) linear gradient, then reverses its direction. The durations of the forward and backward runs are determined by the cell's measurement of the gradient, rather than the gradient's true value. The values of the discrete gradient measurements, $\hat{c}_{1,i}$, sampled over a time $T$, are assumed to be normally distributed with mean $(\partial/\partial t + \mathbf{v} \cdot \nabla ) C(\mathbf{x},t)$ and SD $\sigma = \Pi \times \sigma^{\text{min}}$ (see Eq.~(\ref{minstd})). In the absence of adaptation, this formulation allows one to calculate the expected length of a bacterial run, composed of multiple independent gradient measurements, $\hat{c}_{1,i}$ (Eq.~(S21)). From this, the drift velocity can be written as an infinite series involving the mean reversal rate for cells (Eq.~(S22)). The simplified model successfully predicts the spatiotemporal response of bacteria in experiments (Fig.~\ref{FIG3}C) and could therefore be embedded in advection-diffusion models for chemotaxis \cite{Keller1970,Taylor2012}.

\vspace{-5pt}
\subsection*{Chemotactic Precision Governs Nutrient Uptake}

To investigate the influence of measurement noise on the nutrient exposure across the bacterial population, we used the computational model to analyze the uptake dynamics for different values of $\Pi$. In doing so, we identify ecological tradeoffs that may give rise to the specific precision factor, $\Pi$, exhibited by {\it V. ordalii}. We characterize the potential uptake for each bacterium with position $\mathbf{x}_i(t)$, as the time-dependent concentration of glutamate to which it was exposed, $U_i(t) = C(\mathbf{x}_i(t),t)$. We note that the population-average value of $U$ will in general depend on the size of the domain. However, as 1.5\,mm is beyond the maximum radius for which cells can detect the gradient (see Figs.~\ref{FIG2} and \ref{FIG3}), we fix this value throughout the simulations. Figure~\ref{FIG4}A-C illustrates the simulated trajectories of bacteria for three different values of $\Pi$. The circular shaded regions (blue) represent a measured drift velocity of $|v_{\text{drift}}| > 1 \, \mu$m/s over the first 60\,s after pulse release. The outer radii of these zones for $\Pi=1$ and $\Pi=6.6$ are comparable. SI Appendix Videos S1-S3 show the full response of cells in Fig.~\ref{FIG4}A-C. 

Figure~\ref{FIG4}D demonstrates that the time-dependent population-averaged uptake, $U$, depends strongly on the chemotactic precision factor, $\Pi$. Within $\sim$100\,s following pulse release, the value of $U$ peaks and then slowly declines. When considering the time-averaged potential uptake (over 300\,s) as a function of $\Pi$, for different quintiles of the bacterial population, we find that the nutrient exposure for cells in the top 20\% (in terms of the time-averaged potential uptake) is unaffected by changes in $\Pi$, for $\Pi \lesssim 6$ (Fig.~\ref{FIG4}E). Conversely, the middle to lower quintiles benefit most from changes in $\Pi$, exhibiting a substantial increases in $U$ from $\Pi=100$ to $\Pi=1$.

Bacteria with a precision factor on the order of that observed experimentally ($\Pi = 6.6$) attain a nutrient uptake which is commensurate with the theoretical maximum (Fig.~\ref{FIG4}F). A cell's potential uptake, $U$, depends strongly on its initial position, and Fig.~\ref{FIG4}F shows the time-averaged uptake as a function of initial radial distance, $r_{\text{init}}$, from the pulse center. The results are independent of $\Pi$ for cells that are either very close to, or very far from, the pulse. Beyond $r_{\text{init}} \sim 1500 \, \mu$m, all curves go to zero, and this radius therefore represents the extent of the phycosphere for this pulse size \cite{Hein2016,Smriga2016,Seymour2017}. The confidence intervals corresponding to $\Pi=1$ and $\Pi=6.6$ overlap, indicating that potential uptake values are similar for cells with the theoretical optimal precision and cells with the level of precision determined from experimental data.

It is instructive to consider the relative advantage attained by chemotactic cells compared to non-chemotactic `{\it mutants}', which represent the limiting case where no gradients are detectable. The number of moles of glutamate released in the pulse is given by $n= \pi r^2 h C_0$ where $r=100\, \mu$m and $h=200\, \mu$m are the radius and height of the columnar release, respectively. It is equivalent to study pulse size in terms of $C_0$ or $n$ (horizontal axes of Fig.~\ref{FIG4}G). The relative enhancement of the time- and population-averaged potential uptake due to chemotaxis depends strongly on initial glutamate pulse concentration $C_0$ (Fig.~\ref{FIG4}G). Even for the smallest pulse size studied experimentally ($C_0=1.4\, \mu$M), cells with $\Pi=6.6$ exhibit a $\sim$20\% advantage over non-chemotactic cells. This advantage grows with $C_0$, as more of the chemotactic cells are able to exploit the dynamic hotspot. For $C_0=35\, \mu$M, cells with $\Pi=6.6$ exhibit an 89\% increase in potential uptake over non-chemotactic cells. 

\begin{figure}[h!]
\begin{center}
\includegraphics[width=0.95\columnwidth]{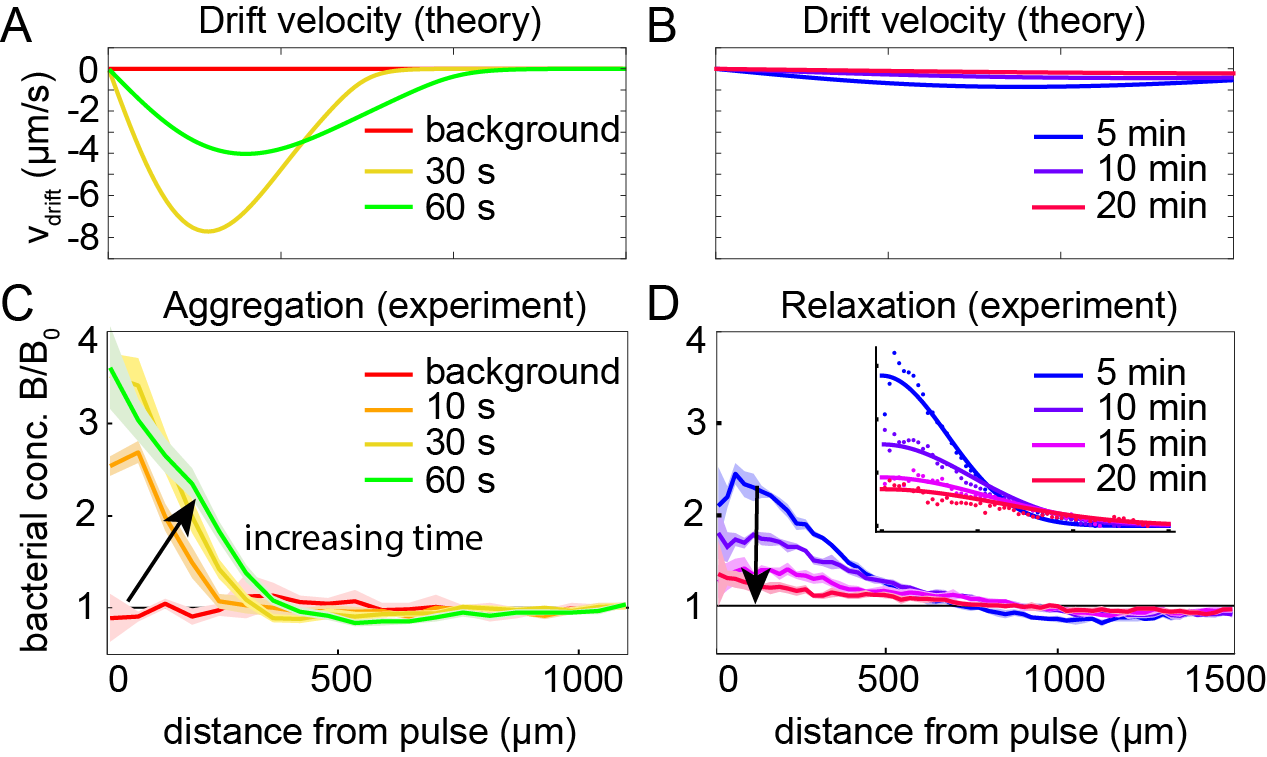}
\caption{Time-dependent bacterial response to photoreleased nutrients of initial concentration $C_0=35\,\mu$M (mass 0.22 pmol). (A,B) Theoretically predicted radial velocity profiles at various times, obtained using Eq.~(S22). (C) Experimental snapshots of bacterial concentration at $t = 10,\ 30,\ 60\,$s after creation of a diffusing nutrient pulse. Conservation of bacteria dictates that beyond the aggregation ($B>B_0$) exists a depletion zone, where $B<B_0$. (D) The bacterial concentration takes approximately 20\,min to relax to $B(r,t) \approx B_0$. Inset shows that $B(r,t)$ for $t \gtrsim 300\,$s is well captured by diffusive spreading alone with a diffusion coefficient $D_B = 165 \, \mu$m$^2$s$^{-1}$. \vspace{-15pt}}
\label{FIG5}
\end{center}
\end{figure}

\vspace{-10pt}
\subsection*{DOM Hotspots as Ecological Units}

These results allow us to quantify the spatial extent and lifespan of bacterial aggregations in realistic environments. One example are phycospheres in the ocean, the regions surrounding individual phytoplankton cells that are rich in dissolved organic matter (DOM) \cite{Smriga2016,Seymour2017}. The continuum theory predicts a short period of active bacterial recruitment via chemotaxis (Fig.~\ref{FIG5}A) followed by many minutes of random motility-induced motion with $v_{\text{drift}}\approx 0$ (Fig.~\ref{FIG5}B). These predictions are consistent with the azimuthally averaged bacterial concentration from experiments (Fig.~\ref{FIG5}C,D). The peak value in bacterial concentration occurs in the center of the glutamate pulse at $t \approx 60\,$s (Fig.~\ref{FIG5}C). The relaxation back to a spatially uniform distribution is well-described by a diffusive process, with an estimated diffusion coefficient of the bacteria of $D_B = 165 \, \mu$m$^2$s$^{-1}$ (inset of Fig.~\ref{FIG5}D). The lifetime of the bacterial accummulation is almost an order of magnitude greater than the duration of chemotaxis, indicating the long-term legacy of short-term initial recruitment. Processes such as collective nutrient cycling \cite{Azam2007, Smriga2016}, horizontal gene transfer \cite{Barlow2009} and infection by pathogens \cite{Moor2017}, which depend on local bacterial concentration, may therefore be influenced well beyond the time when chemotaxis ceases.

\vspace{-5pt}
\subsection*{Discussion and Conclusions} Bacteria in the ocean encounter nutrient pulses from a range of sources. Our experimental system is designed to mimic a range of unsteady nutrient sources, such as the diffusive spreading of a plume behind a sedimenting particle \cite{Stocker:2008fk}, a nutrient filament produced by turbulent mixing \cite{Taylor2012}, or the spreading source from a lysing phytoplankton cell \cite{Barbara2003,Smriga2016}. Through experiment, theory and numerical simulations, we have shown that the chemotactic motion of cells towards an unsteady nutrient source can only occur in discrete zones where the gradient signal is not obscured by noise. Importantly, our results demonstrate that there is a clear and predictable delineation between zones where chemotaxis can occur, and those where it cannot.

For many real nutrient sources, the stochasticity in the chemoattractant is essential in understanding the chemotactic footprint. However, in many existing model for chemotaxis, it is assumed that cells are able to perfectly measure changes in their surrounding chemical concentration. This deterministic sensing is equivalent to setting $\Pi=0$ in our model and assumes variations in the gradient estimate are negligible. The fit of simulations performed in the absence of noise (i.e., $\Pi = 0$) to drift velocity data has a mean fitting error ten times that of the model with $\Pi = 6.6$ (error $5.2\, \mu$m/s compared to $0.51\, \mu$m/s). However, the importance of noise will diminish at higher chemoattractant gradients. As the initial concentration of glutamate in the patch, $C_0$, is increased, the maximum signal bacteria experience, $\max (\partial C / \partial t + \mathbf{v} \cdot \nabla C)$, increases proportionally with $C_0$, while the uncertainty in the measurement scales sublinearly, $\sigma_C \sim \sqrt{C_0}$ (Eq.~(\ref{minstd})). This could explain why models with deterministic sensing \cite{Brown1974, Jackson1987} are capable of matching experiments performed with steep gradients at higher concentrations.

Is it surprising that {\it V. ordalii} operates so close to the limit of sensory precision? On one hand, it is intuitive that more precise sensing should facilitate better responses to the environment, and so natural selection should lead to bacteria capable of making precise measurements of the gradient. There is only a marginal difference between the nutrient exposure for cells with the sensory precision of {\it V. ordalii}, compared to cells with the theoretical optimum sensory precision (Fig.~\ref{FIG4}). However, sensory precision comes at a high cost \cite{Lan2012}. Suppressing internal noise in biochemical networks, for example, generally requires that a cell produce and maintain a greatly increased number of signalling molecules \cite{Lestas2010, Govern2014}, implying that cells must trade-off the costs and benefits of noise suppression. Two potential ways for a cell to increase its chemotaxis performance are (1) by increasing swimming speed \cite{Hein2016}, which comes at the cost of devoting more energy to locomotion \cite{Magariyama1995}, and (2) by increasing its chemotactic precision (i.e., decreasing $\Pi$), which comes at the cost of tighter regulation of noise in the signal transduction pathway \cite{Lestas2010, Govern2014}. Although we cannot here quantify the costs of noise suppression in the chemotaxis pathway, the plateaus in Fig.~\ref{FIG4}E for $\Pi \lesssim 6$ hint towards an optimal value of precision, beyond which the additional resource gain may not outweigh the cost required to suppress internal noise.

We note that noise in the chemotaxis pathway, however, does not always degrade the ability of cells to climb gradients. In contrast with our result on the negative effect of (upstream) counting noise on chemotactic performance, recent studies with {\it E. coli} \cite{Flores2012,Sneddon2012, He2016} demonstrated that (downstream) signaling noise plays an important role in coordinating multiple motors \cite{Hu2013} and can increase chemotactic sensitivity. Rigorous calculations in such model species where the chemotaxis pathway is better established may help shed light on the importance of correlations in measurement noise, out-of-equilibrium dynamics, and simultaneous contribution of multiple noise sources.
 
We have shown that sensory noise places fundamental constraints on the chemotactic abilities of cells and governs the density, spatial extent, and lifespan of bacterial aggregations. The timescale for initial recruitment through chemotaxis (tens of seconds) is much shorter than the lifetime of the bacterial aggregation (tens of minutes). This further highlights the ecological significance of chemotactic navigation during the initial seconds following the occurrence of a pulse, and therefore the crucial role of noise suppression.

From a modeling perspective, the ability to partition complex nutrient landscapes into discrete zones of active chemotaxis will facilitate the conceptual scaling-up from single hotspots to larger domains of an ecosystem, such as the intricate turbulence-induced network of DOM in the ocean. Beyond marine bacteria, the approach of studying chemotactic zones with respect to the underlying gradient signal to noise ratio is expected to find great utility in assessing the performance of other microbes, which have evolved in chemical microenvironments with fundamentally different spatiotemporal properties. 

\matmethods{A detailed discussion of the experimental protocols, numerical simulations and mathematical theory is included in the SI Appendix.

}

\vspace{-10pt}

\acknow{The authors thank V. Sourjik, N. Wingreen, T. Emonet, F. Menolascina, K. Son, V. Fernandez and J. Keegstra for useful discussions. This work was supported by an ARC Discovery Early Career Researcher Award DE180100911 (D.R.B.), The University of Melbourne Computational Biology Research Initiative and HPC system (D.R.B.), a Swiss National Science Foundation Early Mobility Postdoctoral Fellowship (F.C.), a James S. McDonnell Foundation Fellowship (A.M.H.), Army Research Office Grants W911NG-11-1-0385 and W911NF-14-1-0431 (S.A.L.), Simons Foundation Grant 395890 (S.A.L.), a Gordon and Betty Moore Marine Microbial Initiative Investigator Award GBMF3783 (R.S.) and a Simons Foundation Grant 542395 (R.S.) as part of the Principles of Microbial Ecosystems Collaborative (PriME).}
\showacknow

\end{document}